\documentclass[aps,preprint,showpacs,superscriptaddress,groupedaddress]{revtex4-1}

\usepackage[utf8]{inputenc}
\usepackage[french,english]{babel}
\usepackage[T1]{fontenc}
\usepackage[version=3]{mhchem} 
\usepackage{graphicx}
\usepackage{dcolumn}
\usepackage{bm}
\usepackage{xcolor}
\usepackage{caption}

\usepackage{amssymb}
\usepackage{amsmath}
\usepackage{epsfig}
\usepackage[colorlinks=false, pdfborder={0 0 0}]{hyperref}
\usepackage{cleveref}
\usepackage{siunitx}
\usepackage{tikz}
\usepackage{pgfplots}

\usepackage{upgreek}
\usepackage{siunitx}

\usepackage{cleveref}
\crefname{figure}{Fig.}{Figs.}
\crefname{equation}{Eq.}{Eqs.}
\crefname{table}{Table}{Tables}

\begin{document}
\title{Linearity of charge measurement in laser filaments}

\author{Denis Mongin}
\author{Elise Schubert}
\author{Lorena de la Cruz}
\author{Nicolas Berti}
\author{J\'{e}r\^{o}me Kasparian}\email{jerome.kasparian@unige.ch} 
\author{Jean-Pierre Wolf}

\affiliation{Universit\'e de Gen\`eve, GAP, Chemin de Pinchat 22, CH-1211 Geneva 4, Switzerland}

\begin{abstract}
We evaluate the linearity of three electric measurement techniques of the initial electron density in laser filaments by comparing their results for a pair of filaments and for the sum of each individual filament. 
The conductivity measured between two plane electrodes in a longitudinal configuration is linear within \SI{2}{\%} provided the electric field is kept below 100~kV/m. Furthermore, simulations show that the signal behaves like the amount of generated free electrons.
The slow ionic current measured with plane electrodes in a parallel configuration is representative of the ionic charge available in the filament, after several~\si{\micro\s}, when the free electrons have recombined. It is linear within \SI{2}{\%} with the amount of ions and is insensitive to misalignment.
Finally, the fast polarization signal in the same configuration deviates from linearity by up to \SI{80}{\%} and can only be considered as a semi-qualitative indication of the presence of charges, e.g., to characterize the filament length.
\end{abstract}

\maketitle

\section{Introduction}

Ultrashort laser pulses can produce long plasma channels, or filaments~\cite{BraunKLDSM1995,ChinHLLTABKKS2005,CouaiM2007,BergeSNKW2007}, that stem from a dynamic balance between Kerr self-focusing and defocusing by higher-order non-linear effects including ionization. The plasma channel left behind by the filaments is essential for applications such as lightning control~\cite{ZhaoDWE1995,ComtoCDGJJKFMMPRVCMPBG2000,KaspaAAMMPRSSYMSWW2008a}, weather modulation \cite{KaspaAAMMPRSSYMSWW2008a,HeninPRSHNVPSKWWW2011,Ju2014}, or THz generation~\cite{TzortMPAPFMMGBE2002}. Furthermore, the amount and distribution of free electrons in the filament trail is a key ingredient to assess for the self-guiding mechanism, as well as to estimate the losses in view of long-distance propagation~\cite{LaVJCCDJKP1999,RodriBMKYSSSELHSWW2004,Panagiotopoulos2015,Houard2016}.

Several methods have been proposed to characterize the ionization along the filaments, providing a relative longitudinal profile of free charges. These include the observation of nitrogen fluorescence~\cite{HosseLFLARC2003}, the measurement of the laser-generated shockwave~\cite{YuMKSGFBW2003}, holography~\cite{Abdollahpour2011,rodriguez_-line_2008,centurion_holographic_2004},  shadowgraphy~\cite{TzortPFM2000,MinarGCTPDD2009,LiuLLXLMZ2010}, interferometry~\cite{point_two-color_2014,chen_direct_2010}, microwave waveguiding \cite{papeer_temporal_2013,alshershby_diagnosis_2013}, terahertz scattering \cite{bodrov_plasma_2011}, and simple electrical setups measuring conductivity~\cite{tzortzakis_formation_1999,schillinger_electrical_1999,TzortPFM2000,lu_quasi-steady-state_2015,vujicic_low-density_2005}, capacity change due to the fast plasma polarization~\cite{SchilS1999,HeninPKKW2009a,Abdollahpour2011,ionin_plasma_2014,ladouceur_electrical_2001,AkturZFCM2009}, or the ionic current~\cite{Polynkin2012,mongin_conductivity_2016}. While these methods yield qualitatively consistent results~\cite{HosseYLC2004}, their accuracy is still lacking a precise evaluation.

Among all approches, electric measurements are widely used because their experimental implementation is rather straightforward. Furthermore, they appear as the most direct measurement of electron density. However, their results are influenced by screening due to free charges, induction, as well as the plasma dynamics (electron attachment, ion-ion and electron-ion recombination...) that affect the fraction of the charges that reach the electrodes~\cite{ZhaoDWE1995,VidalCCDLJKMPR2000}. Several electrode configurations and signal analysis strategies are promoted by different groups, with an ongoing debate about their precision~\cite{Abdollahpour2011,Polynkin2012}.

Here, we investigate the validity of three commonly used electrode configurations and signal analyses by assessing their linearity with regard to the amount of initial electrons released between the electrodes in the measurement region. Due to non-linearity and intensity clamping in the filaments, this cannot be performed by varying the input pulse energy. Furthermore, in the lack of a widely-accepted technique providing an absolute initial electron density calibration, we used a self-referenced approach requiring no calibration: we compare the signal measured by the setups from two independent filaments with the sum of each filaments individually  (section \ref{sec:longitudinal} and \ref{sec:parallel}). In section \ref{simus} we measure the dependence of the different signals with the input beam energy and compare it with numerical simulations, in order to attribute the measurement technique to either electronic or ionic signal.

\section{Experimental setup}

\begin{figure}[ht]
\begin{center}
\includegraphics[width=12cm, keepaspectratio]{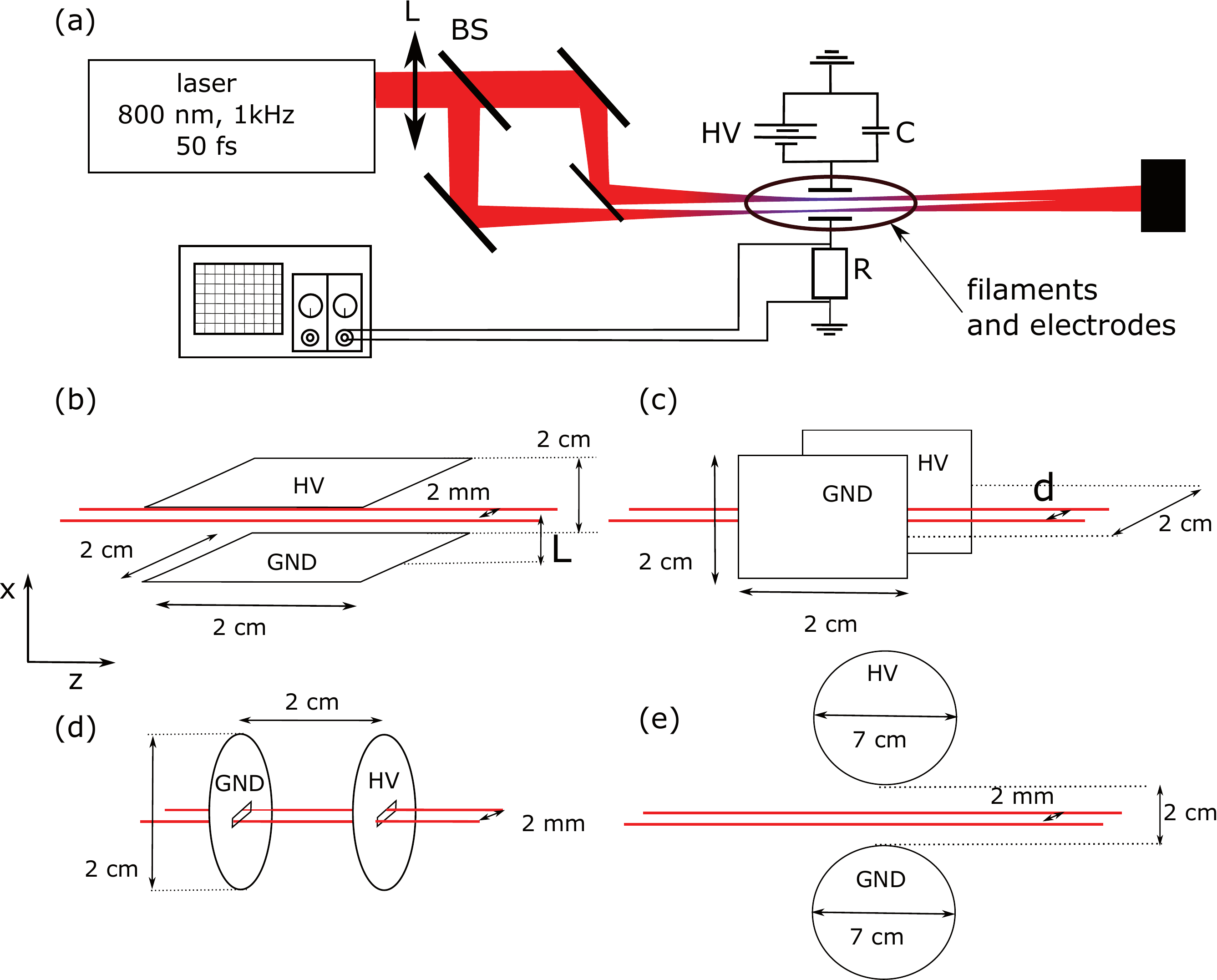}
\end{center}
\caption{Experimental setup. (a) General layout. L:~$f$~=~2~m lens; BS: 70/30 beam splitter; HV: high voltage supply; C: capacitor; R: resistor. Main configurations of electrodes used: (b) parallel configuration, (c) transverse configuration, (d) longitudinal configuration and (e) parallel configuration with spherical electrodes.}
\label{fig:setup}
\end{figure}

The principle of the experimental setup is sketched in \Cref{fig:setup}(a). 
A laser system producing 50 fs pulses at a central wavelength of 800 nm at a repetition rate of 1~kHz  was slightly focused ($f$~=~2~m) and separated on a 70/30 beam splitter into two sub-beams labelled A and B, and  carrying 1.5 and 3~mJ, respectively.
The two beams propagated in air, parallel and at a distance of 2~mm from each other. Each beam generated a single filament. Two plane electrodes made of copper PCB, with a separation of 2~cm were positioned close to the most intense part of the beam, characterized by the strongest visible fluorescence and the maximum signal of the electric measurement. Unless otherwise specified, the filaments propagated in the middle of the gap between the two electrodes.

In a first configuration, hereafter denoted the parallel configuration [\Cref{fig:setup}(b)], 2~cm$\times$2~cm square electrodes were positioned on both sides of the filaments and parallel to the plane containing the two sub-beams. It was used to measure both the fast polarization signal, and the ionic current.

In a second configuration (longitudinal configuration, \Cref{fig:setup}(d)), circular electrodes of 2~cm diameter were placed longitudinally in the propagation direction, with the filament propagating through them, so as to measure the electronic current.

The electrodes were connected to a DC high voltage generator delivering 2 to 10 kV. The transient current generated by the filament is recorded via an oscilloscope (coupling impedance 1~M$\Omega$, capacitance 13~pF) on a resistor $R$~=~47~k$\Omega$ for the parallel configuration (resp. $R$~=~13~k$\Omega$ for the longitudinal), providing a typical temporal resolution of \SI{0.6}{\micro\s} (resp. \SI{0.17}{\micro\s} for the longitudinal configuration). The capacitor load under 10~kV is \SI{1.6e6}{C}, 5 orders of magnitude larger than the total charge collected on the electrodes, ensuring that the charge supply by the capacitor and the associated voltage drop are not perturbing the measurement. The curves were averaged over 500 laser shots.
 
We investigated the linearity of each measurement with regard to both the bias electric field and the initial amount of electrons produced by the laser in the electrode gap. This was achieved respectively by varying the voltage applied to the electrodes and by comparing the sum of the individual signals obtained from the filaments A and B individually (hereafter denoted "A + B") with the signal obtained when both filaments propagated together between the electrodes (hereafter denoted "A and B"). The filaments propagate independently because the $\simeq 30 $~ps time delay between both laser pulses is larger than their duration; However, this delay is shorter than the plasma dynamics so that the subsequent evolution of the charges in both filaments can be considered simultaneous.

 \section{Results and discussion}
 
\begin{figure}[t!]
\begin{center}
\includegraphics[width=1\columnwidth]{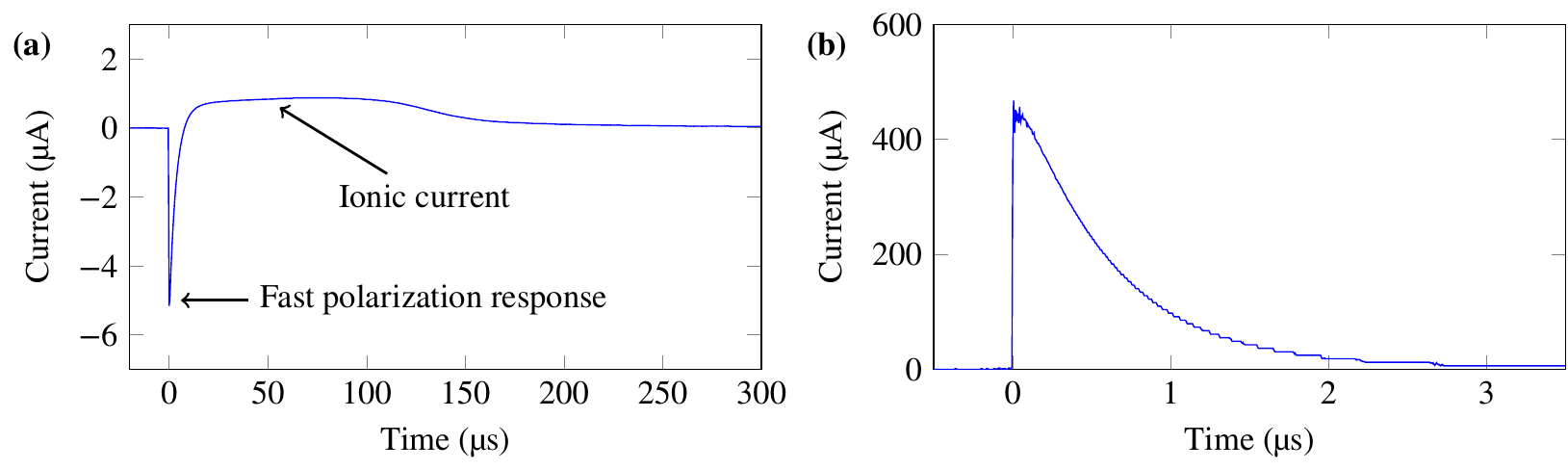}
 	\end{center}
\caption{Typical shape of the raw signal. (a) parallel configuration [\Cref{fig:setup}(b)]; (b) longitudinal configuration (electronic current measurement, \Cref{fig:setup}(d)} 
 	\label{exemple} 
 \end{figure}

As displayed in \Cref{exemple}(a), the typical current recorded on the ground electrode in the parallel configuration exhibits two time scales. 
First, a short negative peak results from the fast polarization of the plasma channel. It is due to charge separation under the action of the external field and induces a transient negative current $J_P = {\partial P}/{\partial t}$,  $P$ being the polarization. This signal is expected to decay within typically 10~ns, mainly governed by the electron attachment to O$_2$ molecules~\cite{ZhaoDWE1995,VidalCCDLJKMPR2000}. Such dynamics is not resolved temporally by our detection setup.
No current due to electron flow is measured since the electron attachment is faster than their transit to the positive electrode. More specifically, electric fields of 1 -- 5$\times 10^5$~V/m yield an electronic mobility of 0.14 -- 0.07~m$^2$/V$\cdot$s~\cite{ZhaoDWE1995}, so that the electrons need 270 -- 700 ns to travel 1~cm to the electrode. After such transit time, an initial electronic density of 10$^{20}$~m$^{-3}$ will have dropped to 10$^{11}$ -- 10$^{8}$~m$^{-3}$, so that the corresponding current is below the nanoampere.

On a longer time scale, the signal exhibits a positive ionic current from \SIrange{20}{400}{\micro\s} (depending on the electric field value) corresponds to the ions travelling to the electrodes~\cite{Polynkin2012}.
Such ionic current can be detected thanks to the slow decay of the ions, as compared to the electrons.
The former is dominated by the recombination of positive ions with negative ions produced by the above mentioned attachment, that is a second-order process. After typically 10~ns, electron-ion attachment is negligible and the density of ions $n_\textrm{ions}$ evolves as:
\begin{equation}
\frac{\mathrm{d}n_{\text{ions}}}{\mathrm{d}t} = - \beta n_{\text{ions}}^2
\end{equation}
$\beta \simeq 2.10^{-13}$~m$^3$/s being the ion-ion recombination rate at 300~K \cite{ZhaoDWE1995}. As a result, for an initial ion density $n_0$
\begin{equation}
n_\textrm{ions} = \frac{n_0}{1+n_0 \beta t} \approx \frac{1}{\beta t}
\label{eq:second_order}
\end{equation}
which is independent from $n_0$ as soon as $n_0 \beta t \gg 1$ so that the ionic current will be essentially driven by the volume of the filament where this condition is fullfilled. The ion density is still as high as \SI{5e16}{m^{-3}} after \SI{100}{\micro\s}, allowing the detection of the corresponding current in spite of  the low ionic mobility of typically  \SI{2.5e-4}{m^2/Vs}~\cite{ZhaoDWE1995}, which results in a velocity in the 10~m$\cdot$s$^{-1}$ range.

A typical signal in the longitudinal configuration is displayed on \Cref{exemple}(b). The  positive signal is mainly due to the flow of electrons  between the electrodes through the filament.
No positive signal is observed at long times, due to a more efficient ion neutralization than in the parallel configuration: the  positive and negative ions travel along the same pathway and in opposite directions, increasing the probability for charges of opposite polarities to meet and neutralize.

\subsection{Parallel electrode configuration }\label{sec:parallel}
\subsubsection{Ionic current}

\begin{figure}[t]
\begin{center}
 \includegraphics[width=0.4\columnwidth]{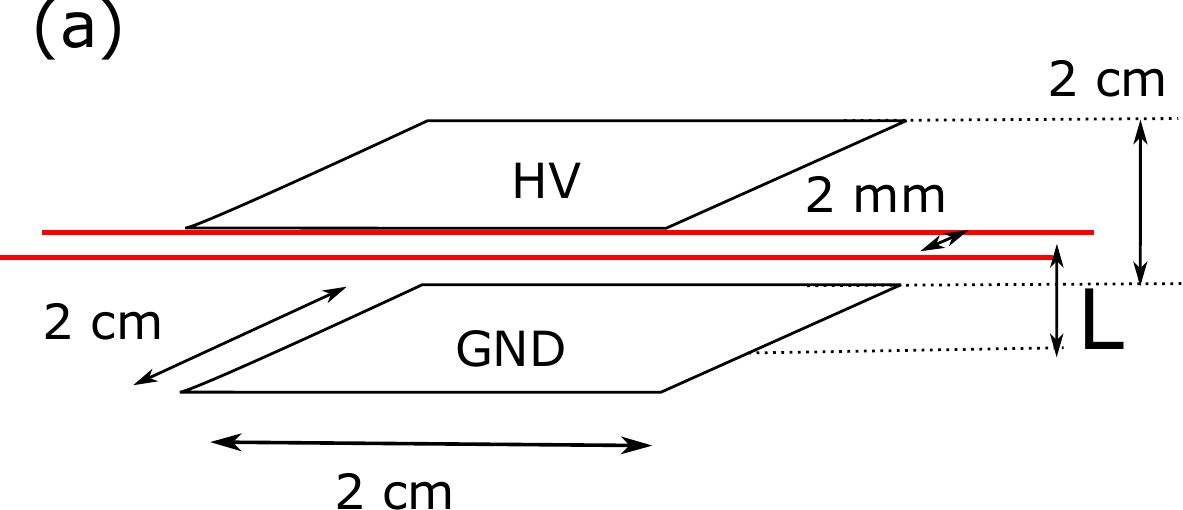} \\
\includegraphics[width=1\columnwidth]{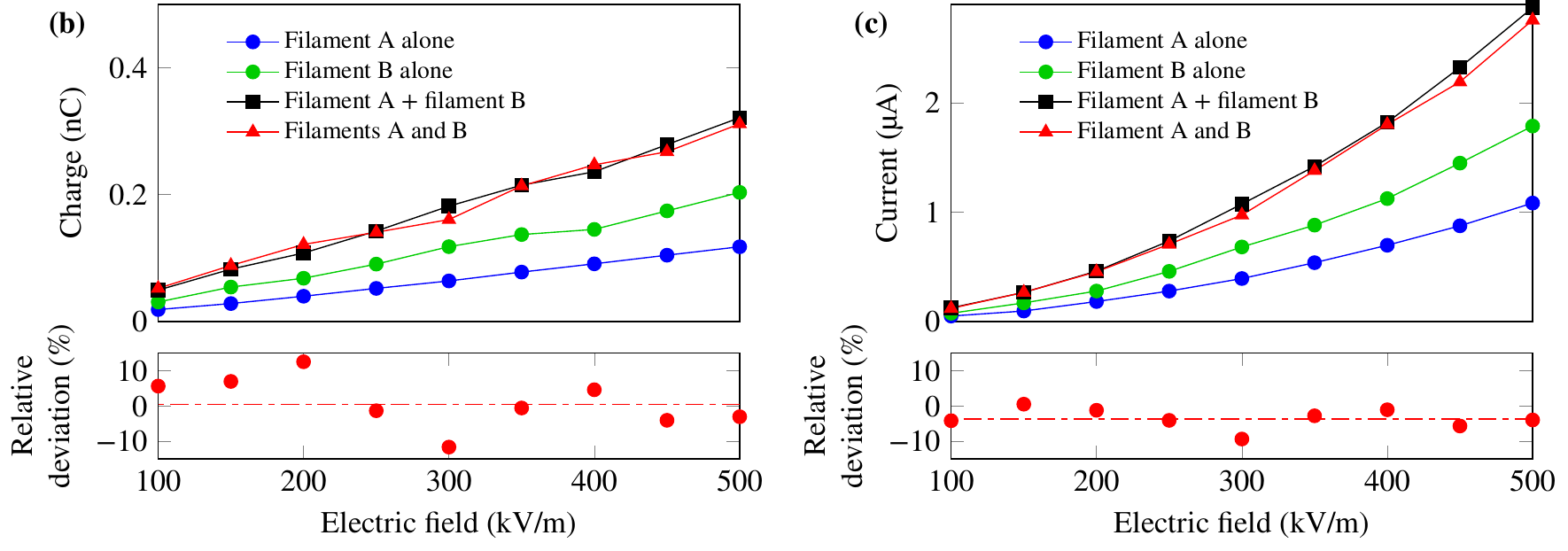}
 	\end{center}
\caption{(a) Configuration for the measurement of the (b) total collected charge and (c) peak ionic current.}
 		\label{fig:parallele} 
 \end{figure}

In the case of the ionic current, the sum of the charges collected from the individual filaments is equal to that of the filaments measured together, within a relative deviation of 12\% at most  [\Cref{fig:parallele}(b)].  We can thus conclude that the measurement of the collected charges in this configuration is linear with respect to the amount of electrons initially produced by the filaments.

The peak current is also linear with respect to the initial amount of electron produced by the filament, with a maximum deviation of  10~\% [\Cref{fig:parallele}(c)]). Furthermore, using 7 cm diameter spherical electrodes spaced by 2~cm[\cref{fig:setup}(e)] leads to similar results as the plane electrodes (\Cref{fig:parallele}(b) and (c)) alongside with a reduced deviation to 2\% because of the higher homogeneity of the electric field.

Approaching the filaments from either electrode temporally splits the ionic current peak into two sub-peaks corresponding to the respective times of flight of the positive and negative ions to their respective target electrode. 
The maximum current of the second sub-peak keeps linear with the initial electron density of the filament within 10~\% even when the filaments are as close as 2~mm to either electrode. This robustness with regard to static misalignments is remarkable, considering the finite size of the electrodes and the associated inhomogeneities in the electric field. 
Conversely, the maximum of the first sub-peak (i.e charges having a shorter path to travel) deviates from linearity by up to 40\% and tends to deform when the filaments approach one of the electrodes. Such perturbations are due to the influence of the fast negative polarization signal, as well as to a contribution of electrons reaching the electrodes at a short distance.

It can be observed in \Cref{fig:parallele}(b) that the total charge collected on the electrodes increases linearly with the electric field [\Cref{fig:parallele}(b)], while the peak current rises quadratically [\Cref{fig:parallele}(c)]. This behavior can be understood by considering \Cref{eq:second_order}, with a time of arrival $t_\textrm{f} \simeq (\mu_\textrm{ion} E)^{-1}$, $\mu_\textrm{ion} \simeq$ \SI{2.5e-4}{m^2/Vs} being the ionic  mobility~\cite{ZhaoDWE1995}. The ion density reaching the electrodes is therefore $n_\mathrm{ions}=\mu_\textrm{ion} E / \beta$, proportional to the applied electric field. Furthermore, the current is proportional to both the ion density and their velocity $\mu_\textrm{ion} E$, so that the current is proportional to $\left(\mu_\textrm{ion} E\right)^2 / \beta$.
 
\begin{figure}[h!]
\begin{center}
$\vcenter{\hbox{\includegraphics[width=0.4\columnwidth]{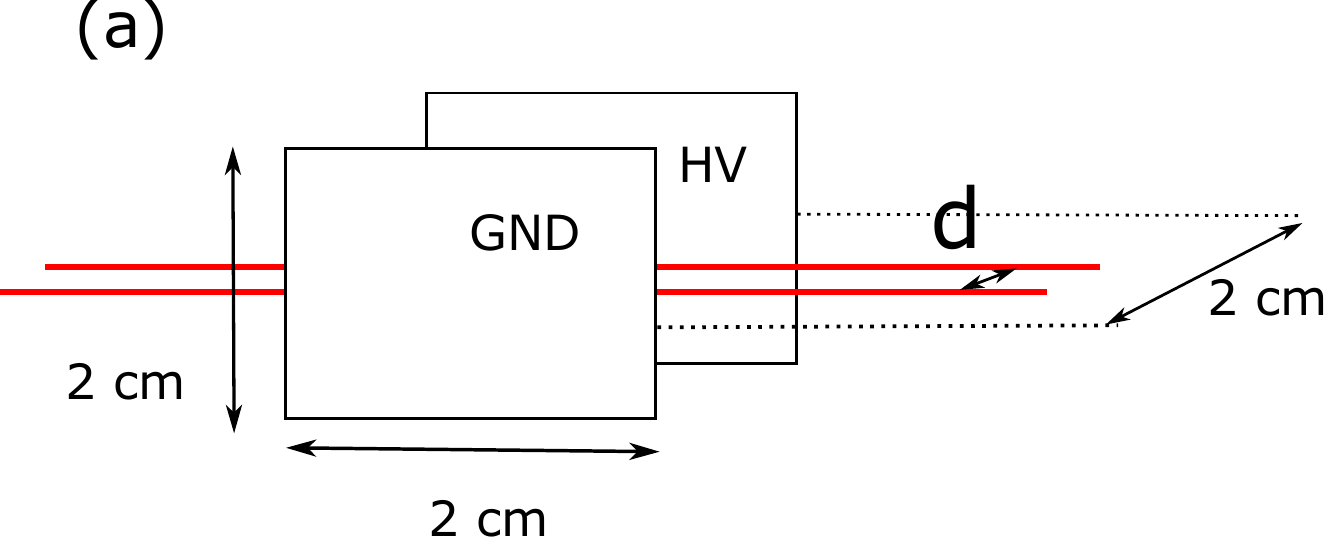}}}$
 $\vcenter{\hbox{\includegraphics[width=0.55\columnwidth]{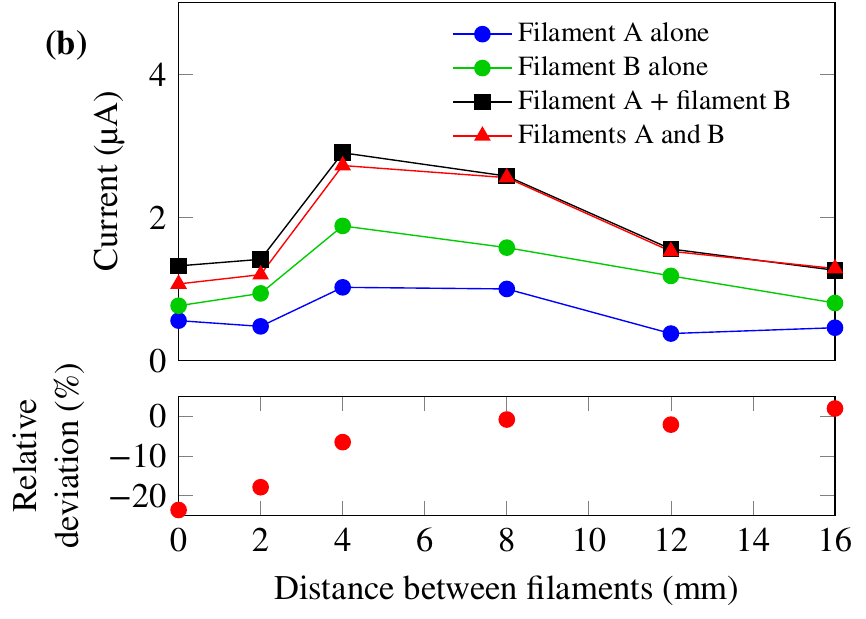}}}$

 	\end{center}
\caption{(a) Transverse configuration; (b) Effect of the distance between the filaments on the ionic current for an electric field of \SI{5e5}{kV/m}}
 		\label{bosse_perp} 
 \end{figure}

We also investigated its sensitivity to the screening of the electric field by space charges, by switching from the parallel configuration [\Cref{fig:setup}(b)] to a configuration where the electrodes are perpendicular to the plane containing the two filaments (transverse configuration, see \Cref{fig:setup}(c) and \Cref{bosse_perp}(a)). In this configuration, the charges released by each filament screen the other filament from the opposite electrode. The measurement significantly deviates from linearity if filaments are closer than 4~mm to each other [\Cref{bosse_perp}(b)]. The deviation is negative, i.e. the measurement underestimates the ion density by 20\% when both filaments are measured together (A and B), evidencing a measurable although limited screening. In contrast, for longer separation distances, the relative deviation to linearity keeps below 10\%, comparable to the parallel configuration. 

In summary, both the peak ionic current and the corresponding total collected charge are representative of the ion density in the plasma channel after \SIrange{10}{100}{\micro\s}, except in the case of multiple filamentation, where the typical distance between filaments is in the millimeter-range~\cite{PetitHNBJKBSSRKSWW2011,Ettoumi2015b,mongin2016gas}.

\subsubsection{Fast electronic polarization signal}

\begin{figure}[h!]
\begin{center}
$\vcenter{\hbox{\includegraphics[width=0.4\columnwidth]{schema_transverse.pdf}}}$
 $\vcenter{\hbox{\includegraphics[width=0.55\columnwidth]{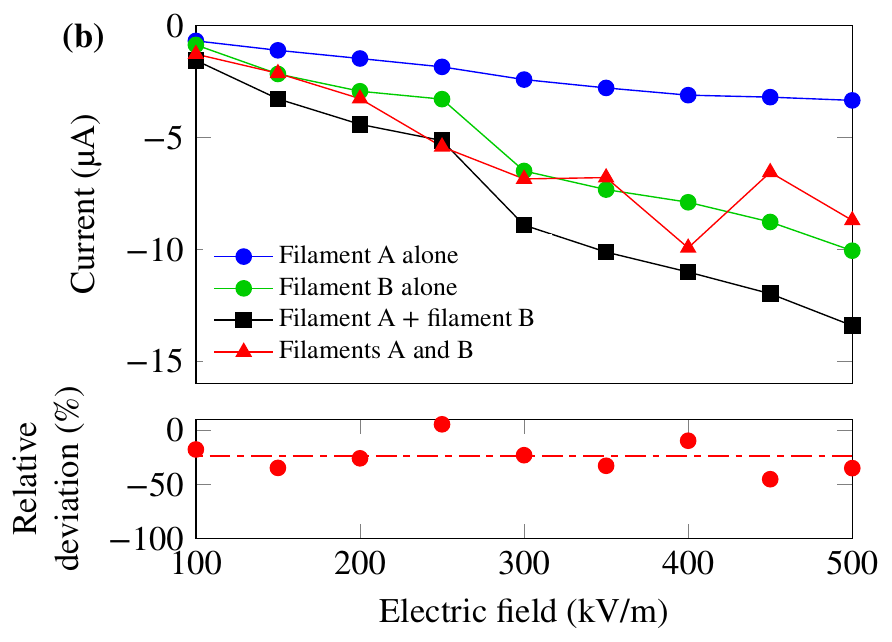}}}$
	\end{center}
\caption{(a) Parallel configuration for the measurement of the fast polarization signal and (b) its magnitude as a function of the applied voltage.} 
	\label{court_para} 
\end{figure}

The amplitude of the peak associated with the fast polarization signal in the parallel configuration is quite linear with the applied electric field. In contrast, it deviates from linearity with regard to the the initial amount of electrons between the electrodes  by up to 54\% [\Cref{court_para}(b)] (in other words, the sum of the measurements of individual filament differs from the measurement of the two filaments together by up to 54\%, the average of this discrepancy being 24\%).

This strong deviation can be explained by considering the mutual interaction between the two filaments considered as lines of dipoles. 
More specifically, the electric field $E$ displaces the free electrons from their parent ion by  $x_\textrm{e} = \mu_\textrm{e} E t$, $\mu_\textrm{e}$ being the electronic mobility~\cite{ZhaoDWE1995} and $t$ the time elapsed since the ionization. Consequently, each filament with a cross-section $a$ and a free electron density $n_\textrm{e}$ has a dipole moment per unit length ${\textrm{d}p}/{\textrm{d}z} = a e n_\textrm{e} x_\textrm{e}$. It induces an electric field on the second filament, located at the same height $x$ and at a transverse distance $d$. The electric field $E_\textrm{ind}$ induced at a longitudinal position $z_0$ of the second filament can be calculated by integrating the field induced by the infinitesimal dipole moment 
${\textrm{d}p}/{\textrm{d}z}$, over the length $l$~=~2~cm of the first filament:

\begin{equation}
E_{\textrm{ind}}(z_0,d,t) = -\frac{a e n_\textrm{e}(t) \mu_\textrm{e} E t}{4 \pi \varepsilon_0}\int_{-z_0}^{l-z_0} {\frac{\textrm{d}z}{\left( d^2+z^2 \right)^{3/2}}}
\label{eq:2f_Edipz}
\end{equation}

In turn, this induced electric field will affect the filament polarization. We numerically calculated the latter as a function of time based on \Cref{eq:2f_Edipz}, relying on the temporal evolution of the electron density $n_\textrm{e}(t)$ provided by the plasma evolution model described in \cite{Schubert2016}.
We obtain a reduction of the electronic polarization signal by 24\% for a peak electron density $n_{\mathrm{e},0}$~=~\SI{3e21}{m^{-3}}, in line with values reported in filaments~\cite{BergeSNKW2007}.

Finally, the measurement and its linearity are highly sensitive to the distance between the filaments and the electrodes. 
When the filaments get closer than 5~mm to one electrode (either HV or ground), the polarity of the signal even reverses. This inversion is due to the positive contribution resulting from the collection of an electronic current when the filaments are too close to one electrode. In an electric field of 5$\times$10$^5$~V/m the electrons with a mobility of 7.6$\times$10$^{-2}$~m$^2$/Vs~\cite{ZhaoDWE1995} need 130~ns to reach the electrode 5~mm away. Their density at arrival is then of the order of 10$^{16}$~m$^{-3}$, five order of magnitude larger than when the filaments are 1~cm away from the electrodes.

\subsection{Longitudinal electrode configuration}\label{sec:longitudinal}
\subsubsection{Electronic current}
\begin{figure}[h!]
\begin{center}
$\vcenter{\hbox{\includegraphics[width=0.4\columnwidth]{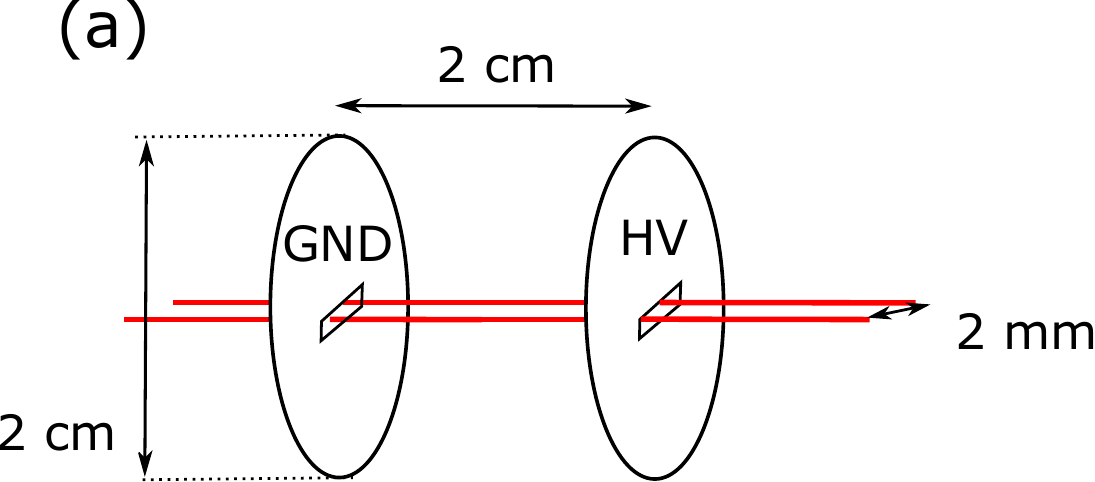}}}$
 $\vcenter{\hbox{\includegraphics[width=0.55\columnwidth]{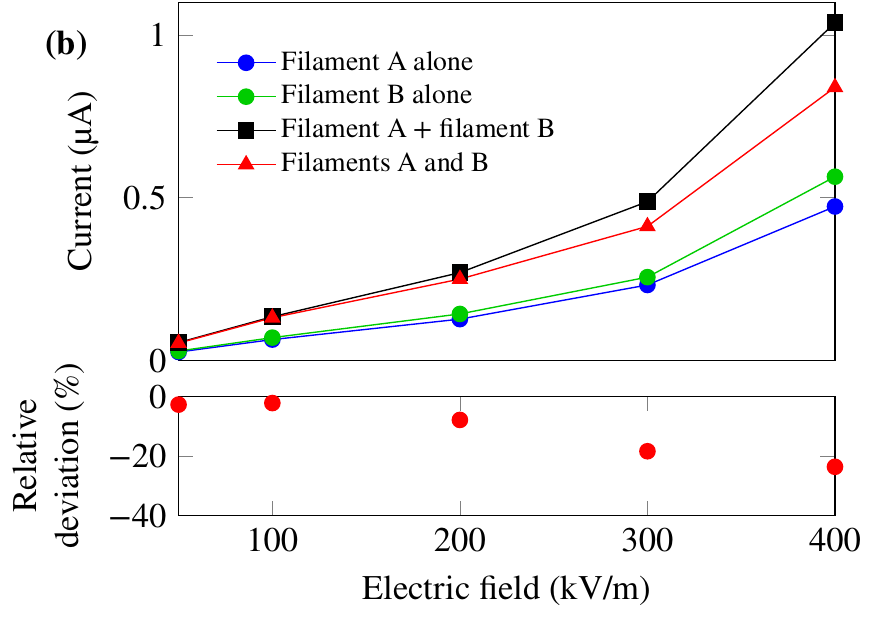}}}$
	\end{center}
\caption{(a) Longitudinal configuration and (b) peak electronic current as function of the applied external field in this configuration.} 
	\label{court_long}
\end{figure}

In the longitudinal configuration, where the filaments propagate through the electrodes without touching them [\Cref{fig:setup}(d) and \Cref{court_long}(a)], the peak current is very linear (within 2\%) with the initial amount of electrons released between the electrodes, provided the applied electric field stays below 100~kV/m [\Cref{court_long}(b)]. Conversely, it deviates by up to 20\% for 400~kV/m. The total collected charge behaves similarly, showing that this approach could be relevant to estimate the electron density and its evolution along the filament.

\subsection{Comparison with a filament propagation model}\label{simus}

In order to relate more precisely the measurements in each configuration with the charges in the filaments, we compared them with the production of free electron, and of ions after electron attachment, simulated in a filamentation model.

\begin{figure}[t]
\begin{center}
\includegraphics[width=1\columnwidth]{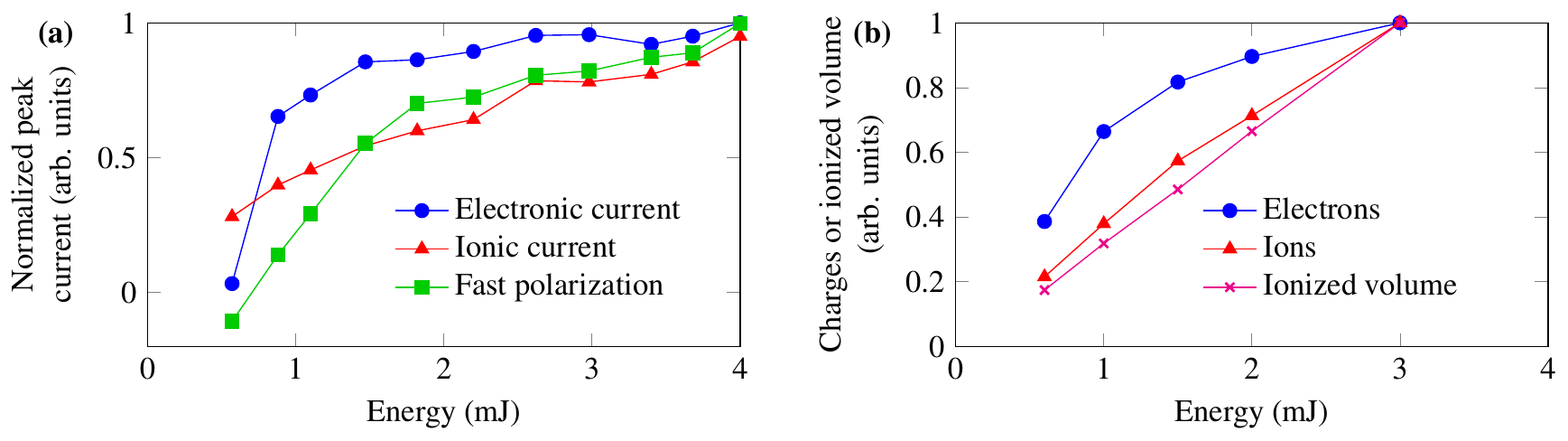}
\end{center}
\caption{(a) Energy dependence of the electric measurement techniques of filament plasma: electronic current in the longitudinal configuration, ionic current and fast polarization signal in the parallel configuration. (b) Simulated  energy dependence of the electrons and ions amount produced by a filament between the electrodes, as well as the ion density after \SI{100}{\micro s} ionized volume.}
\label{comparaison_signaux}
\end{figure}

Experimentally, the incident pulse energy was varied from \SIrange{0.5}{4}{mJ}, and we checked that a single filament was generated in all cases.
Each measurement was performed at the propagation distance corresponding to the maximum signal. The ionic current and the fast polarisation signal were measured in the parallel configuration for an electric field of 300~kV/m. The electronic current was measured in the longitudinal configuration with an electric field of 50~kV/m, accordingly with the limitation evidenced above. The results are displayed in \Cref{comparaison_signaux}(a).

These measurements were compared with the simulated amount of initial charges of the filament, i.e., the electron (resp. ion, at the time when they reach the electrodes) density integrated both across the filament cross-section, and longitudinally over the region of the filament that is located between the electrodes. The propagation dynamics was calculated using the model based on the unidirectional pulse propagation equation (UPPE)~\cite{KolesMM2002} and described in~\cite{Berti2015} except for the ionization model, that is based on the PPT formalism~\cite{BergeSNKW2007}. The subsequent evolution of the charges relied on the plasma model of~\cite{Schubert2016}. The results are displayed in \Cref{comparaison_signaux}(b) together with the volume where the ion density exceeds \SI{1e18}{m^{-3}}, i.e., where the conditions of \Cref{eq:second_order} are fullfilled after \SI{10}{\micro\s}.

The experimental electronic current saturates when the incident energy increases [\Cref{comparaison_signaux}(a)]. As confirmed by the simulation (\Cref{comparaison_signaux}(b), blue curve), this saturation corresponds to intensity clamping. The further slow increase in the signal is due to the rise in intensity in the close vicinity of the filament, leading to a slight increase of the total charge, and therefore of the current.

In contrast, the ionic current increases linearly with the beam power [\Cref{comparaison_signaux}(a)], like the simulated total ion number after \SI{100}{\micro\s}, i.e. when they typically reach the electrodes [\Cref{comparaison_signaux}(b)]. As discussed above (See \Cref{eq:second_order}), the ion density at that time is almost independent from their initial density as long as the latter is above 10$^{18}$~m$^{-3}$. The total ion number and the associated ionic current are therefore governed by the volume where this condition is fulfilled. This volume increases linearly with the incident energy ("ionized volume" in \Cref{comparaison_signaux}(b)), and is 10 times larger than that of the filament core where the free electron density is sufficient to effectively contribute to the electron measurement (n$_e \sim 5.10^{21}$ m$^{-3}$). Furthermore, the simulations show that the latter depends much less on the pulse energy than the former. 

Finally, the fast polarization signal has an intermediate behavior between the two previous curves, which suggests that both electrons and ions contribute to this signal.

\section{Conclusion}
In conclusion, three variants of electric measurement techniques of the initial electronic density in laser filaments were investigated and compared. Two configurations out of three have been validated as they offer a good linearity with respect to the initial amount of electrons released in the laser beam.

The longitudinal configuration yields an acceptable linearity when the applied electric field is kept below \SI{100}{\kV\per\m}, with a deviation lower than  \SI{5}{\%}. It shows a saturation behavior with the increase of the beam energy due to intensity clamping in the filament. It therefore seems to be the best candidate for a measurement that intends to measure the initial free electron density in the filament. 

On the other hand, measuring either the peak current or the total collected charge in the long positive swing (\SI{\sim 100}{\us}) on the electrodes parallel to the beam yields a signal insensitive to misalignments, as already put forward by~\cite{Polynkin2012}. This signal is quite linear with the ionic charge in the filament but is difficult to link directly to the initial electron density in the filament as the ion and electron decay kinetics are very different. This signal should be considered as an indicator of the ionic charge available in the filament, that is the charge after most of the electrons have recombined or attached to O$_2$, rather than as a measurement of the initial plasma free electrons. It is interesting to note that the timescale of this signal is in the microsecond range, i.e. the timescale of high-voltage triggering~\cite{ComtoCDGJJKFMMPRVCMPBG2000}. 

Finally, the fast polarization signal leads to relative deviations on the linearity of initial electon density up to \SI{54}{\%}, consistent with the measurements of~\cite{Abdollahpour2011}. This is therefore not a accurate measurement when one aims to investigate the electron density in a filament but it is still a semi-qualitative indication of the presence of charges, hence a valid way to characterize the filament length.

This study should contribute to the assessment of ionization measurement techniques in the context of laser filamentation and help experimentalists choose the most relevant approach to perform accurate relative measurements of initial charges in laser-induced plasma channels. 

 \section*{Funding}
 European Research Council, Advanced Grant ``Filatmo''.

\section*{Acknowledgments}
Technical support by M. Moret was highly appreciated.

\end{document}